\documentclass{article}
\usepackage[utf8]{inputenc}
\usepackage{graphicx}
\usepackage{amsmath}
\usepackage{amssymb}
\usepackage[T1]{fontenc}
\usepackage{authblk}

\title{Overlapping domains for topology optimization of large-area metasurfaces}
\author[1]{Zin Lin\thanks{zinlin@mit.edu}}
\author[1]{Steven G. Johnson}
\affil[1]{Department of Mathematics, Massachusetts Institute of Technology, Cambridge MA 02138, USA}

\date{June 2019}

\begin{document}

\maketitle

\begin{abstract}
We introduce an overlapping-domain approach to large-area metasurface design, in which each simulated domain consists of a unit cell and overlapping regions from the neighboring cells plus PML absorbers. We show that our approach generates greatly improved metalens quality compared to designs produced using a locally periodic approximation, thanks to $\sim 10\times$ better accuracy with similar computational cost. We use the new approach with topology optimization to design large-area ($200\lambda$) high-NA (0.71) multichrome and broadband achromatic lenses with high focusing efficiency ($\sim 50\%$), greatly improving upon previously reported works.
\end{abstract}


\section{Introduction}
Inverse-design techniques~\cite{sell2017large,lin2018topology,pestourie2018inverse,liu2018generative,lin2019topology,chung2019high,phan2019high} have received increasing attention as a powerful approach for the design of optical metasurfaces (surfaces with aperiodic subwavelength patterns designed for free-space wavefront engineering)~\cite{capasso2018future,yu2014flat,aieta2012aberration,aieta2015multiwavelength,khorasaninejad2015achromatic,khorasaninejad2017visible,khorasaninejad2017achromatic,arbabi2017controlling,su2018advances,chen2018broadband,wang2018broadband}. For example, with freeform topology optimization~\cite{jensen2011topology,molesky2018inverse}, the resulting devices showed enhanced functionalities such as angular phase control~\cite{lin2018topology} or large-angle beam deflection~\cite{sell2017large}. However, early designs were limited to small devices, whereas many practical applications require diameters $\gtrsim 1000 \lambda$~\cite{capasso2018future} that would require enormous computational resources for full Maxwell simulations. More recently, new inverse-design techniques have become capable of large-area design, using either an interpolated library of \emph{offline} simulation results for a few parameters~\cite{pestourie2018inverse} or freeform topology optimization (in which every ``pixel'' is a degree of freedom) with \emph{online} Maxwell solvers~\cite{lin2019topology}. 
The fundamental strategy to rapidly model a large-area metasurface is to divide it up into many unit cells and then simulate each cell independently. The set of simulated near-fields gathered from each cell are then ``stitched together'' (using the equivalence principle~\cite{pestourie2018inverse}) to approximate the total electromagnetic field anywhere above the surface. The key question is what to choose for the boundary conditions of these unit cells. One common choice is a Bloch-periodic boundary condition~\cite{yu2014flat,aieta2012aberration,aieta2015multiwavelength,khorasaninejad2015achromatic,khorasaninejad2017visible,khorasaninejad2017achromatic,arbabi2017controlling,su2018advances}, which corresponds to a locally periodic approximation (LPA)~\cite{pestourie2018inverse}. While LPA has proven to work well for monochromatic or small numerical-aperture (NA) metasurface designs\cite{capasso2018future,yu2014flat,aieta2012aberration,aieta2015multiwavelength,khorasaninejad2015achromatic,khorasaninejad2017visible,khorasaninejad2017achromatic,arbabi2017controlling,su2018advances,chen2018broadband,wang2018broadband,pestourie2018inverse}, it might not be suitable for designs which require more rapid variations in the dielectric structure, such as a high-NA multi-chrome metalens~\cite{lin2019topology,chung2019high}. 

\begin{figure}
    \centering
    \includegraphics[scale=0.35]{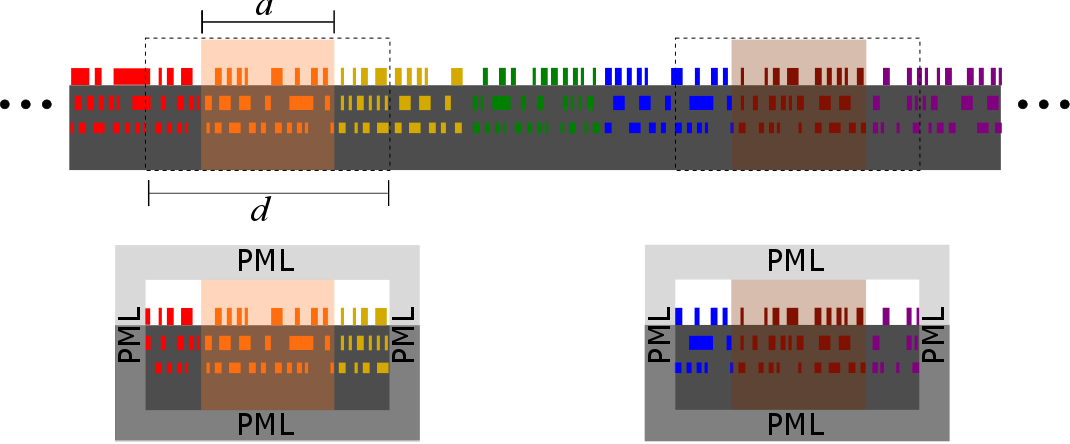}
    \caption{Schematic illustration of the overlapping domain method. An arbitrary aperiodic multi-layered meta-structure (top) is divided into basic unit cells of size $a$ (for example, shaded areas). Each simulation domain (bottom) has a size $d \geq a$ with overlapping regions from the neighboring unit cells and is terminated by perfectly matched layers (PML). The total electric field over the entire metasurface is approximated by the set of near-fields restricted to the corresponding unit cells. }
    \label{fig0}
\end{figure}

Here, we introduce a new approach, called an overlapping-domain approximation (ODA), which is more accurate than LPA for topology-optimized metalens design with similar computational efficiency, and is inspired by domain-decomposition iterative solvers~\cite{gander2019class} but omits the iteration. In this approach, instead of sharply divided boundaries, each simulation domain contains the corresponding unit cell and some spatial overlap regions from the neighboring unit cells.  Fig.~\ref{fig0} shows a schematic of our approach, in which an arbitrary aperiodic metasurface is divided into unit cells of size $a$ (two of which are shaded in Fig.~\ref{fig0}), but the simulated domains have size $d \ge a$ with overlapping regions from the neighboring unit cells. The simulations are additionally padded by absorbing boundaries such as perfectly matched layers (PML)~\cite{taflove13} instead of Bloch-periodic boundaries.  From each simulated domain, we discard the electromagnetic field data in the overlapping regions: the fields in the size-$a$ unit cells are used to construct approximate fields anywhere above the surface similar to previous work~\cite{lin2019topology}. Here, the motivation for PML is that  applications such as lenses and beam-forming typically do not involve long-range light-propagation \emph{within} the surface, so the suppression of such inter-cell scattering by the PML is appropriate (and superior to artificially periodic or Dirichlet boundaries) as long as $d$ is sufficiently large. In fact, we find that the introduction of PML reduces much of the LPA error (an observation also made by a recent work~\cite{phan2019high}) and that the error continues to fall with increasing overlap; overlap is especially crucial for situations with strong near-field inter-cell coupling.  In particular, we find that $d \lesssim 1.5a$ is enough for accurate modeling of many lens designs.  (In contrast, there will be other applications such as grating couplers or excitation of surface resonances~\cite{Perez-Arancibia:18} in which long-range surface propagation is crucial, but such propagation implies a nearly periodic surface in which LPA will be accurate. In this sense, ODA and LPA are complementary.)  Not only do we demonstrate ODA's improved accuracy over LPA (Sec.~2), but we also combine it with topology optimization~\cite{jensen2011topology,molesky2018inverse} to design large-area ($200\lambda$) high-NA (0.71) multi-chrome (red-green-blue/RGB, in Sec.~3) and broadband achromatic lenses (in Sec.~4) with high focusing efficiency ($\approx 50\%$), greatly improving upon reported results in previous works~\cite{pestourie2018inverse,lin2019topology}.

ODA is directly inspired by domain-decomposition techniques such as the Schwarz algorithm~\cite{zepeda2018nested,stolk2013rapidly,chen2013source,gander2019class,vouvakis2004modeling,dolean2009optimized}.  These are iterative methods to find the \emph{exact} solution of a large problem by repeatedly solving small problems in overlapping domains, where the boundary conditions of each domain are updated from its neighbors in the subsequent iteration.   More generally, the solution of subproblems in small domains can be used as a preconditioner for many different iterative methods in linear algebra~\cite{strang2007computational}, and in such contexts some authors have even used overlapping domains with PML~\cite{leng2015overlapping,stolk2017improved}. From this perspective, ODA is simply the initialization step (or the 0th-order iteration) of a Schwarz-like algorithm, but the key observation is that this 0th-order approximation is sufficiently accurate for many metasurface design problems.  Additional Schwarz iterations or similar schemes could be used for arbitrarily improved accuracy (or error estimates), but during metasurface optimization it is more important to solve the problem quickly than it is to solve it exactly, since the solution only needs to be good enough to point the optimization algorithm at an improved design for the next optimization step.  More accurate simulations can be used at the end of the design process for validation; in this paper, we validate using brute-force finite-difference time-domain (FDTD)~\cite{oskooi2010meep} simulations.

\section{Locally periodic vs. overlapping-domain approximations}

\begin{figure}
    \centering
    \includegraphics[scale=0.47]{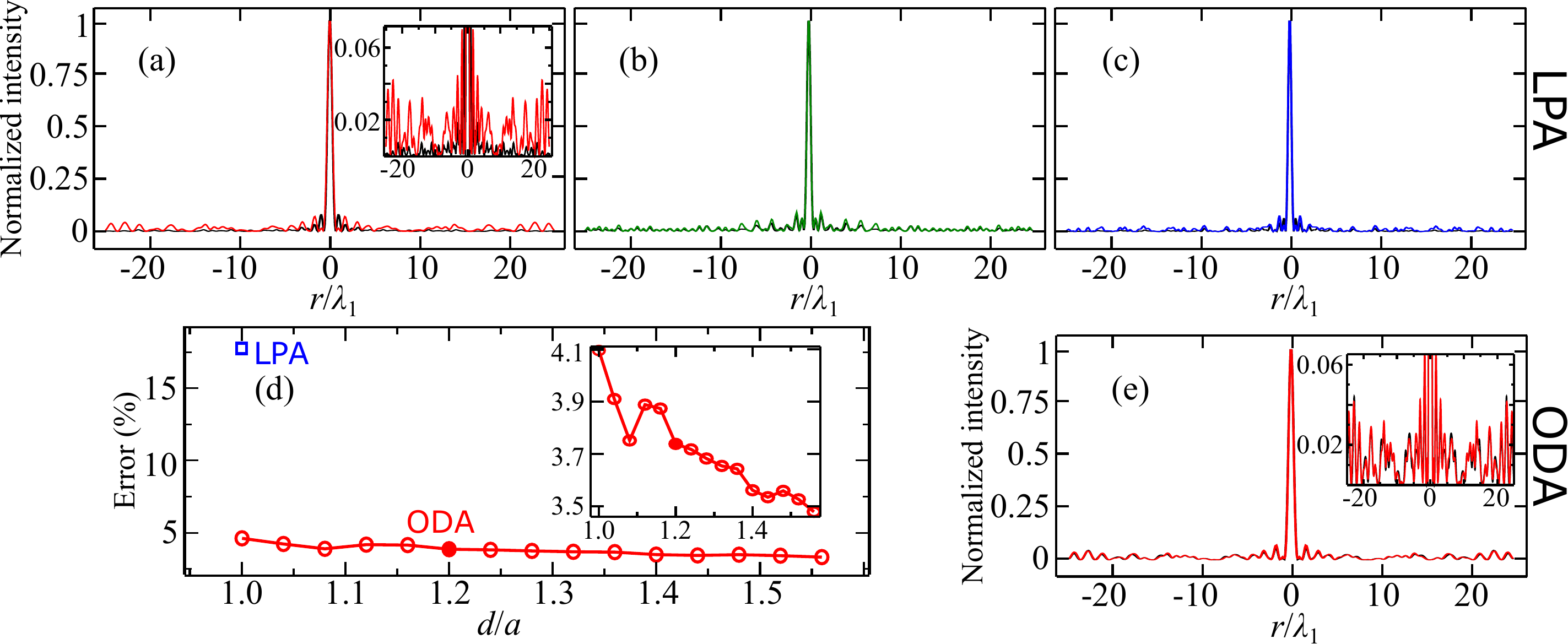}
    \caption{Normalized far-field intensities $I(r)$ of a 2d metalens optimized, under the locally periodic approximation (LPA), for (a) red $\lambda_1=700~\mathrm{nm}$, (b) green $\lambda_2=560~\mathrm{nm}$, and (c) blue $\lambda_3=480~\mathrm{nm}$ wavelengths. The lens is 200$\lambda_1$ wide and has a focal length of 100$\lambda_1$, corresponding to $\mathrm{NA}=0.71$. The intensities are measured at the focal distance. The black lines denote LPA predictions while the colored lines denote rigorous full-wave FDTD simulations of the entire metalens at the corresponding wavelengths. The inset in (a) shows the detailed errors between LPA prediction and FDTD simulation at the red wavelength. (d) Far-field profile prediction errors in percentage for the red wavelength. The error $\Vert I_\text{FDTD}-I_{d/a} \Vert_2 / \Vert I_\text{FDTD} \Vert_2$ is defined in terms of L$_2$ norms for various overlapping domain sizes $d$. The unit-cell size $a$ is $5\lambda_1$. Note the blue square denoting the LPA data point ($\approx 17\%$); the errors are reduced by about $5\times$ for PML-terminated ODA (red circles). Inset: enlarged view of ODA errors.  (e) Comparison of the far-field intensities between the full-wave FDTD simulation and the overlapping domain prediction at $d/a=1.2$, corresponding to the red solid data point in (d).} 
    \label{fig1}
\end{figure}

Fig.~\ref{fig1}(a--c) demonstrates the limitations of LPA for a multi-chrome high-NA metalens design at RGB wavelengths ($\lambda_1=700~\mathrm{nm},~\lambda_2=560~\mathrm{nm},~\lambda_3=480~\mathrm{nm}$). The metalens consists of two layers of TiO$_2$ (refractive index $n=2.35,2.41,2.49$ at RGB), each $280~\mathrm{nm}$ thick and separated by $70~\mathrm{nm}$; the lower layer is buried in silica ($n\sim1.5$). The lens has a width $200\lambda_1$ and an NA of 0.71 and is designed by topology optimization under the locally periodic approximation (LPA) with the basic unit cell size, $a=5\lambda_1$, using the approach described in Ref.~\cite{lin2019topology}. The intensities are measured at the focal distance. The black lines denote LPA predictions while the red, green and blue lines denote rigorous full-wave FDTD simulations~\cite{oskooi2010meep} of the entire metalens at the respective wavelengths, showing significant discrepancies between the two: rigorous simulations exhibit noisy side lobes while LPA predicts a relatively clean Airy~\cite{khorasaninejad2016metalenses} profile. It is worth noting that the overly optimistic results predicted by LPA indicate that topology optimization did manage to produce an optimal design which works well under LPA, but that the design contain rapid structural variations that violate the slowly-varying assumption behind LPA. Since the discrepancies between LPA and FDTD are apparently more pronounced for the red wavelength (Fig.~\ref{fig1}a inset), we proceed to examine this case more closely. In particular, the error in the far-field intensities, defined as $\Vert I_\text{FDTD}-I_{d/a} \Vert_2 / \Vert I_\text{FDTD} \Vert_2$ where $\Vert I \Vert_2 = \sqrt{\int I(x)^2 dx}$ is integrated over the focal plane, is as much as 17\% under LPA ($d/a=1$). In contrast, Fig.~\ref{fig1}(d) shows that the error dramatically falls once we introduce the PML-termination even without any overlap (an observation also made by a recent work~\cite{phan2019high}) and continues to fall with overlapping domains ($d/a > 1$) as described above, and in fact only $d \le 1.5a$ suffices to lower the error by about a factor of 5 relative to LPA. This is also shown by Fig.~\ref{fig1}(e), which reveals that ODA yeilds much better agreement with rigorous FDTD simulations.  We expect that an overlap region will become even more important in metasurfaces with features that touch or cross unit-cell boundaries, especially for metallic structures that might exhibit strong near-field coupling between adjacent unit cells.

The introduction of PML combined with only a small overlap of $d \le 1.5a$ lowers the error to a few percent or less in this structure, which requires only a modest increase in the computational cost of the subdomain simulations. The basic reason for this is that, in a metalens where light propagates mainly through (rather than along) the surface, the biggest effect of neighboring cells arises from the near fields of adjacent structures. We expect that larger overlapping regions may prove beneficial for taking into account stronger inter-cell scattering effects such as those found in higher dielectrics and plasmonic structures. Of course, there are also small corrections due to in-plane scattering from far-away cells, which leads to an error that decays extremely slowly with $d$ for $d \gg a$. If one wished to correct for such small long-range errors, it would be more efficient to perform Schwarz (or similar) domain-decomposition iteration~\cite{gander2019class} rather than to increase the overlap $d$. However, the agreement in Fig.~\ref{fig1}(e) is already much better than has typically been used for metasurface designs in past work.

\section{Topology optimization with overlapping domains}

\begin{figure}
    \centering
    \includegraphics[scale=0.4]{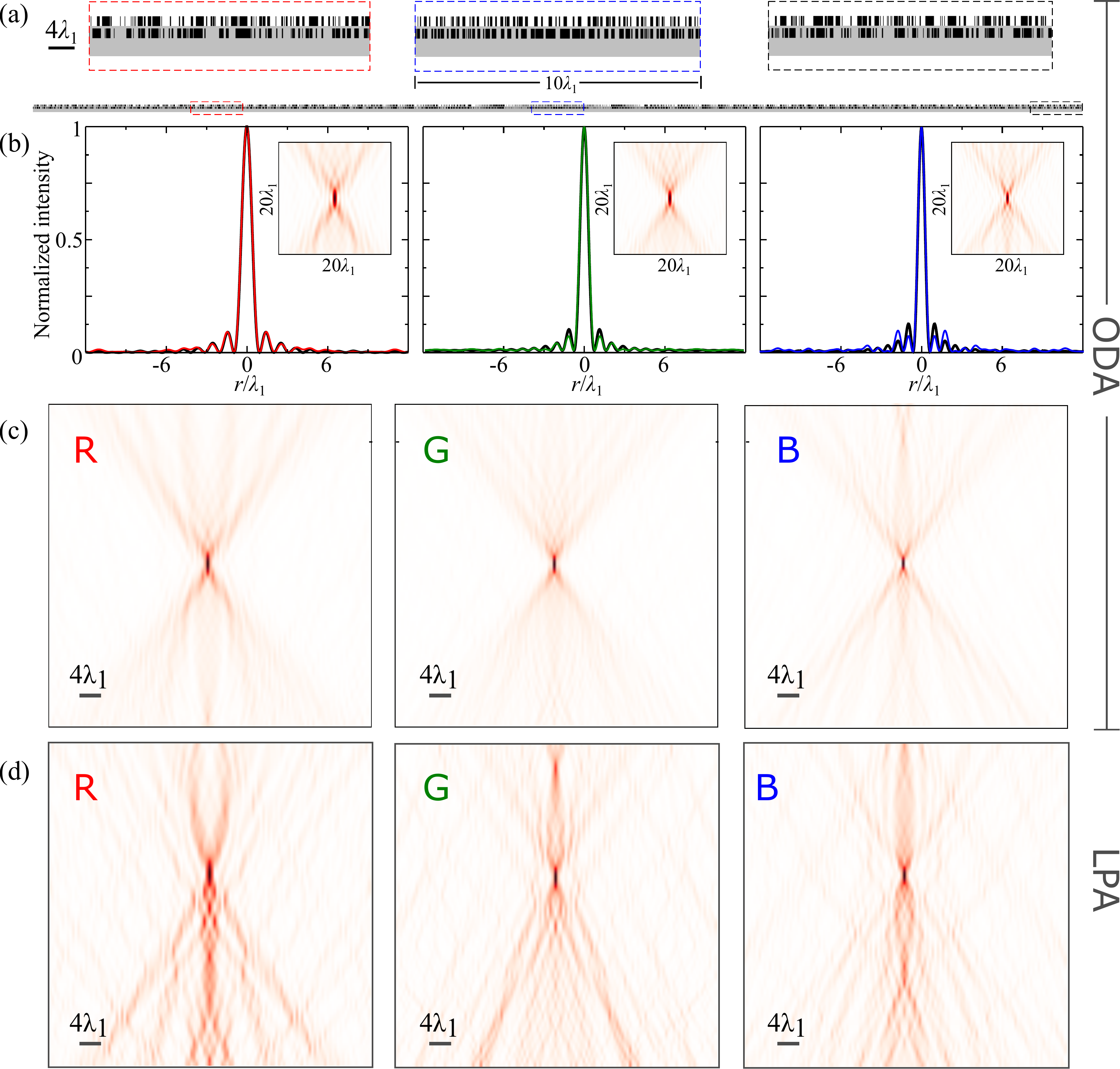}
    \caption{(a) Topology-optimized 2d multi-chrome metalens ($\mathrm{NA}=0.71$) consisting of two layers of TiO$_2$, each $280~\mathrm{nm}$ thick and separated by $70~\mathrm{nm}$. The bottom layer is embedded in the silica substrate. The lens is optimized under the overlapping domain approximation with $a=20\lambda_1,~d/a=1.2$. The total width of the lens is $200\lambda_1$; a few portions of the lens are magnified for easy viewing (note the scale bars). (b) The far-field profile at the focal distance (red, green or blue line) shows good agreement with that of a monochromatic lens at the corresponding wavelength (black solid line). ($\lambda_1=700~\mathrm{nm},~\lambda_2=560~\mathrm{nm},~\lambda_3=480~\mathrm{nm}$) The insets show far-field profiles within $20\lambda_1 \times 20\lambda_1$ windows. (c) The detailed views of the far-field intensity profiles within an enlarged window (size: $50\lambda_1 \times 50\lambda_1$) centered at the focal point. In comparison, the far-field profiles of a metalens designed by LPA (d) exhibit markedly inferior performance such as parasitic side lobes and secondary focal points.  } 
    \label{fig3}
\end{figure}

Encouraged by the error reduction observed in the previous section under ODA, we proceed to perform an ODA-based topology optimization for a metalens with the same dimensions, material parameters and target wavelengths as above. In particular, the optimization involves $10^4$ degrees of freedom (DOF) trying to maximize the minimum of focal intensities~\cite{lin2019topology} at three different wavelengths ($\lambda_1=700~\mathrm{nm},~\lambda_2=560~\mathrm{nm},~\lambda_3=480~\mathrm{nm}$). Fig.~\ref{fig3}(a) shows the binarized~\cite{jensen2011topology} topology-optimized ODA design with $a=20\lambda_1,~d/a=1.2$. The normalized far-field intensity profiles at the focal distance, Fig.~\ref{fig3}(b), show that the metalens (colored lines) is performing virtually as well as a monochromatic lens (black lines) of the same NA at the corresponding wavelength. All three focal spots are diffraction-limited and the focusing efficiencies (defined as the fraction of incident power focused within a radius of 3~FWHM from the focal spot; FWHM~= full width at half-maximum) around the focal spot divided by the total incident flux~\cite{arbabi2015subwavelength}) are found to be 45\%, 48\%, and 35\% respectively. The field profiles in Fig.~\ref{fig3} are computed by full FDTD simulations (ODA is only used during optimization), but we find that ODA produces nearly indistinguishable results ($\lesssim 4\%$ error).  The superior performance of ODA is even more dramatically demonstrated by comparing Fig.~\ref{fig3}(c) to a metalens optimized \emph{using LPA} shown in Fig.~\ref{fig3}(d): the latter (again, in full FDTD simulations performed after optimization) exhibits undesirable noise and parasitic focal points in the far field~\cite{lin2019topology}. To the best of our knowledge, although high-NA achromatic metalenses with much smaller device sizes have been proposed~\cite{shrestha2018broadband}, our design is the first multi-chrome (RGB) high-NA metalens to achieve good focusing performance over a large device size: for comparison, in Table~\ref{tab}, we have compiled a few of the most recent achromatic metalens designs reported to date. As noted in Ref.~\cite{shrestha2018broadband}, it is extremely challenging to achieve a high-efficiency design over a broad bandwidth, a large diameter, and a high numerical aperture all simultaneously; our results show that an appropriately formulated topology-optimization framework combined with a vastly expanded design space can successfully address such issues. Although our proof-of-concept results are in two dimensions, we have shown that similar techniques scale to full 3d designs~\cite{lin2019topology}.

\begin{table}
\centering
\begin{tabular}{|c|c|c|c|c|c|}
\hline
    & NA & Diameter & 2d or 3d & Bandwidth & Efficiency \\
 \hline
 Chen {\it et al.}~\cite{chen2019broadband} & 0.2 & $38\lambda$ & 3d & 460--700$\mathrm{nm}$ & $\sim 30\%$ \\
 Shrestha {\it et al.}~\cite{shrestha2018broadband} & 0.88 & $71\lambda$ & 3d & 1.2--1.4$\mathrm{\mu m}$ & Not reported. \\
 Pestourie {\it et al.}~\cite{pestourie2018inverse} & 0.3 & $361\lambda$ & 2d & RGB & $\sim 30\%$~\cite{Pestourie19:private} \\
 Chung {\it et al.}~\cite{chung2019high} & 0.99 & $18\lambda$ & 2d & 450--700$\mathrm{nm}$ & 27\% \\
 This work (Sec.~3) & 0.71 & $200\lambda$ & 2d & RGB & $\sim 43\%$ \\
 This work (Sec.~4) & 0.71 & $200\lambda$ & 2d & 480--700$\mathrm{nm}$ & $\gtrsim 50 \%$ \\
 \hline
\end{tabular}
\caption{A comparison of achromatic metalens designs. $\lambda$ denotes the longest design wavelength in each reference. \label{tab}}
\end{table}

\begin{figure}
    \centering
    \includegraphics[scale=0.6]{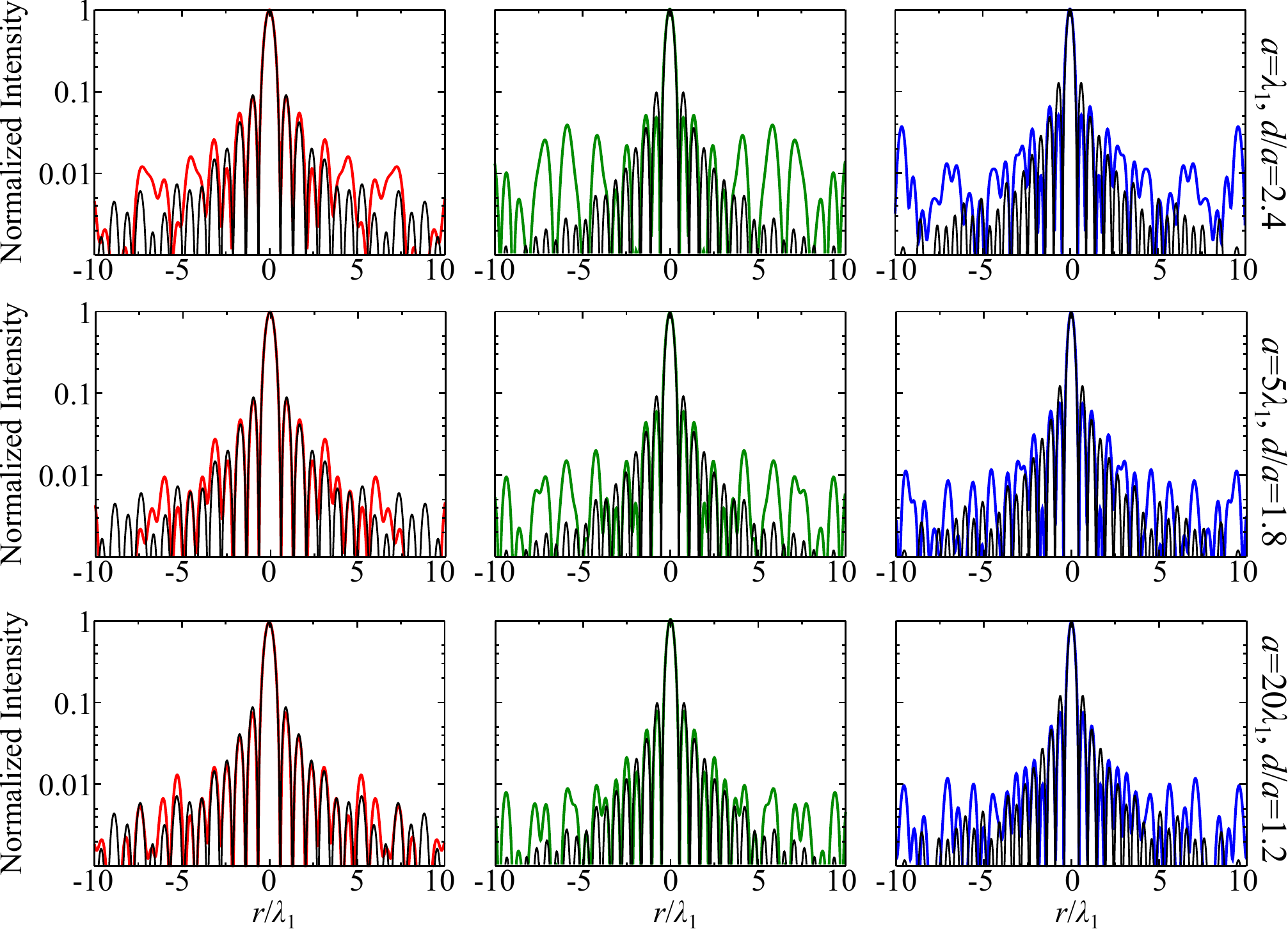}
    \caption{Normalized far-field intensities (logarithmic scale) of three different multi-chrome metalens designs ($\mathrm{NA}=0.71$) optimized under the overlapping domain approximation (ODA) with various unit cell sizes. Upper panel: $a=\lambda_1,~d/a=2.4$; middle panel: $a=5\lambda_1,~d/a=1.8$; lower panel: $a=20\lambda_1,~d/a=1.2$. The black solid lines denote the diffraction-limited far-field profiles obtained from monochromatic designs; the red, green and blue lines denote the full-wave FDTD simulations at the corresponding wavelengths ($\lambda_1=700~\mathrm{nm},~\lambda_2=560~\mathrm{nm},~\lambda_3=480~\mathrm{nm}$). For increasing $a$, the profiles show smaller side noise and approach that of an ideal monochromatic design at the corresponding wavelength. } 
    \label{fig2}
\end{figure}

Here, we would like to emphasize that as in our previous work~\cite{lin2019topology}, we do \emph{not} restrict ourselves to sub-wavelength unit cells, unlike many previous authors~\cite{capasso2018future,yu2011light,yu2014flat,aieta2012aberration,aieta2015multiwavelength,khorasaninejad2015achromatic,khorasaninejad2017visible,khorasaninejad2017achromatic,arbabi2017controlling,su2018advances,chen2018broadband,wang2018broadband}: large unit cells can be thought of in LPA as utilizing many diffraction orders~\cite{pestourie2018inverse,lin2019topology} while ODA allows us to go further and take full advantage of the aperiodic degrees of freedom. To further elucidate the influence of the unit cell size $a$ on ODA-based designs, we perform three independent optimizations with three different $a$ values: small ($a=\lambda_1,~d/a=2.4$), intermediate ($a=5\lambda_1,~d/a=1.8$) and large ($a=20\lambda_1,~d/a=1.2$). While we found that all three cases lead to good designs with diffraction-limited focusing, it is also instructive to carefully examine the noise characteristics for different $a$s. Fig.~\ref{fig2} plots FDTD-simulated focal intensity profiles (red, green, and blue lines) on a logarithmic scale for an easy viewing of the noise floor away from the focal peak; each profile nearly matches that of an ideal monochromatic lens (black lines) at the corresponding wavelength, and also shows excellent agreement with ODA (not shown since indistinguishable from FDTD, errors $\lesssim 4\%$). The largest-$a$ results (lower panel) clearly have the smallest side noise ($<1\%$) for all three wavelengths. 

\section{High-NA large-area broadband achromatic focusing}

\begin{figure}
    \centering
    \includegraphics[scale=0.3]{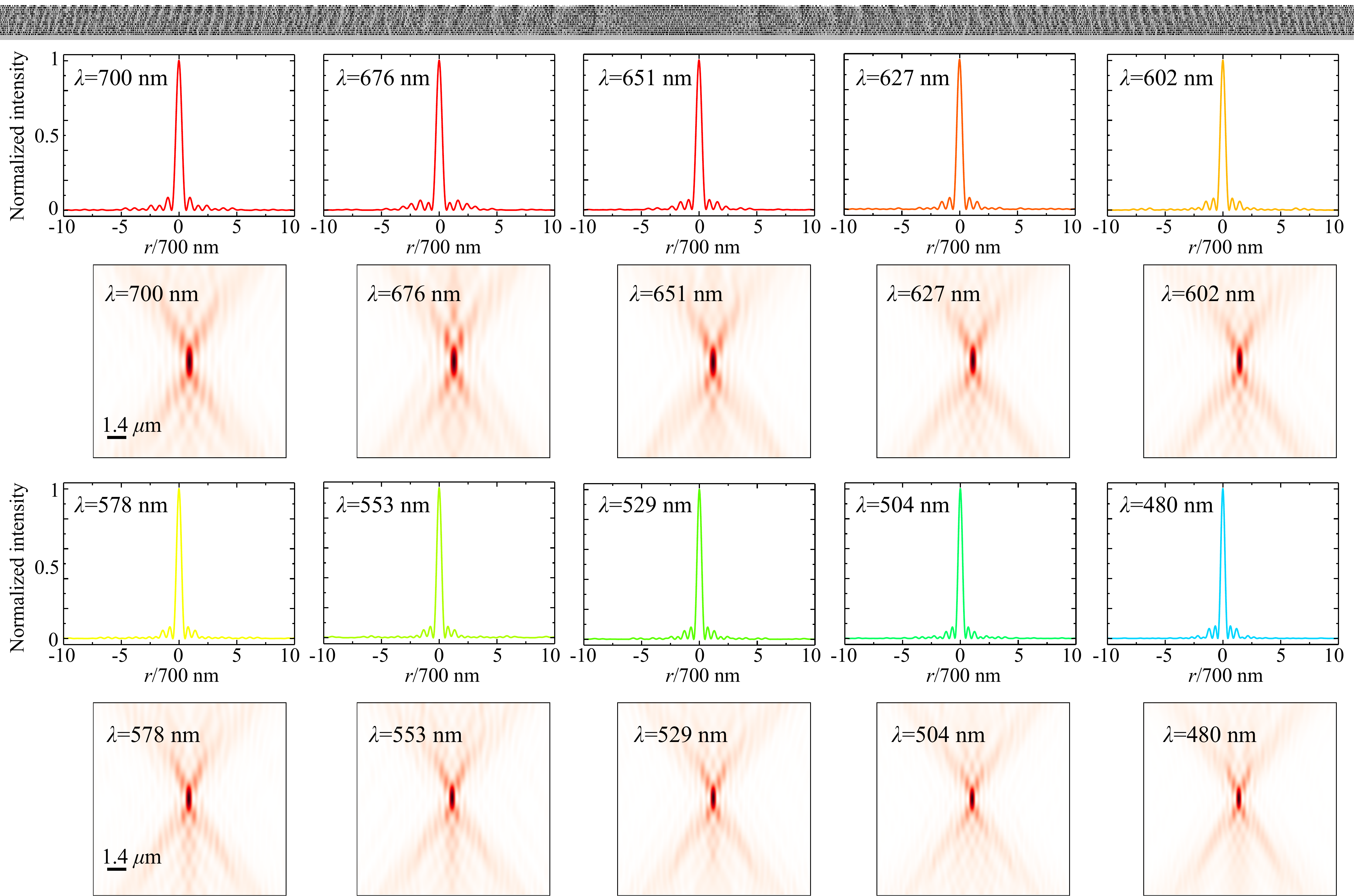}
    \caption{Topology-optimized 2d broadband achromatic metalens ($\mathrm{NA}=0.71$ and $\text{width}=200\lambda$) made up of 15 TiO$_2$ layers. Each layer is $140~\mathrm{nm}$ thick and spaced by $70~\mathrm{nm}$ SiO$_2$.} 
    \label{fig4}
\end{figure}

The primary advantage of topology optimization is that it thrives on an enormous design space.  For only a few degrees of freedom (DOF), hand designs and human intuition can be very powerful. However, given a vast number of DOF and more challenging set of design goals, which simple approaches can no longer handle, topology optimization tends to perform extremely well---an observation known as ``the blessing of dimensionality''~\cite{gershenfeld1999nature}, which may also be related to the recent big-data triumph of deep neural networks over logic-based artificial intelligence~\cite{hinton}. As a testament to this idea, we  present a 15-layer TiO$_2$ metalens with $\mathrm{NA}=0.71$ and $\text{width}=200\lambda$ which involves $1.5 \times 10^5$ DOF trying to maximize the minimum of focal intensities~\cite{lin2019topology} at 10 equidistant wavelengths over the visible spectrum ($480\mathrm{nm}$--$700\mathrm{nm}$). The optimal design (Fig.~\ref{fig4}) clearly exhibits diffraction-limited focusing  with an average focusing efficiency exceeding $50\%$ (in contrast, see Table~\ref{tab}). 

\section{Conclusion and outlook}
In summary, we have introduced an overlapping domain approximation to large-area metasurface design. In this approach, each independently simulated domain consists of a basic unit cell and the overlapping regions from the neighboring cells, and is terminated by PML instead of Bloch boundaries. The approach can be fruitfully considered within the context of iterative domain decomposition techniques such as Schwarz method. If desired, the full Schwarz scheme can be followed to desired accuracy, or a few Schwartz iterations could alternatively provide an estimate of the error introduced by the approximation. We employed the overlapping domain approach to the design of high-NA large-area multi-chrome 2d metalenses, showing \emph{first-in-class} designs that exceed the performance of LPA-designed metalenses. 

In this work, we have presented theoretical designs to demonstrate the power and versatility of our approach. While these designs, with a minimum feature size of $14~\mathrm{nm}$ and a maximum aspect ratio of 20:1, are  extremely challenging to fabricate with the existing technologies, they use physically realistic materials and lengthscales---since the Maxwell equations are scale-invariant~\cite{JoannopoulosJo08-book}, these designs could be fabricated for longer wavelengths.  However, there are well-known regularization procedures to obtain more manufacturable designs from topology optimization that could be applied in future work for practical structures at visible wavelengths~\cite{jensen2011topology,christiansen2015creating}. Additionally, we hope that the possibility of vastly superior performance, which can only be unlocked by multi-layered designs, will help motivate the development of new fabrication technologies. In particular, encouraged by the recent advances in novel nano-scale 3d fabrication techniques~\cite{oran20183d}, we anticipate that topology-optimized 3d designs such as ours will play an increasingly important role in future photonic technologies. We are currently pursuing multi-layered meta-structures via 3d-printing technologies such as Nanoscribe lithography~\cite{buckmann2012tailored}.

\section*{Funding}
This work was supported by the U. S. Army Research Office through the Institute for Soldier Nanotechnologies under grant W911NF-13-D-0001

\end{document}